\documentclass[a4paper,11pt]{amsart}

\usepackage[colorlinks=true,
linkcolor=blue,
citecolor=blue,
filecolor=blue, 
urlcolor=blue   
]{hyperref}

\usepackage{url}

\usepackage{amssymb}
\usepackage{color}

\usepackage{bbm}

\usepackage[normalem]{ulem}
\usepackage[utf8]{inputenc}
\usepackage{graphicx}

\usepackage{soul}

\usepackage{todonotes}

\presetkeys{todonotes}{fancyline, color=green!40, size=\tiny}{}

\usepackage[backend=bibtex]{biblatex}
\bibliography{bib}
\renewbibmacro{in:}{}

\newcommand{\ptclean}[1]{}

\newcommand{\fpsi}{\psi}

\newcommand{\deltazgriem}{h}

\newcommand{\VtwoOrF}{{f}}

 \newcommand{\HTVI}{{H^V_{T,I}}}

\newcommand{\faVI}{{f_{a,I}^V}}
\newcommand{\faSI}{{f_{a,I}^S}}
\newcommand{\fabSI}{{f_{ab,I}^S}}
\newcommand{\HTTI}{{H_{T,I}^T}}
\newcommand{\HTSI}{{H_{T,I}^S}}
\newcommand{\HLSI}{{H_{L,I}^S}}

\newcommand{\lambdarsquare}{\frac{ r^2}{\ell^2}}

\newcommand{\kappahere}{{{}K}}

\newcommand{\mhere}{{}\mu}
\newcommand{\ourellnormal}{{}l}
\newcommand{\efindex}{{}I}
\newcommand{\modeindex}{\efindex}
\newcommand{\Nman}{{}^{n}N}

\newcommand{\ourntwo}{{n-1}}

\newcommand{\bean}{\begin{eqnarray}\nn}

\newcommand{\nn}{\nonumber}

\newcommand{\zgriem}{{ \mathring{\mathfrak g}}}

\DeclareFontFamily{OT1}{rsfs}{}
\DeclareFontShape{OT1}{rsfs}{m}{n}{ <-7> rsfs5 <7-10> rsfs7 <10->
rsfs10}{} \DeclareMathAlphabet{\mathscr}{OT1}{rsfs}{m}{n}

\newcommand{\eq}[1]{\eqref{#1}}

\newcommand{\bel}[1]{\begin{equation}\label{#1}}
\newcommand{\beal}[1]{\begin{eqnarray}\label{#1}}
\newcommand{\beadl}[1]{\begin{deqarr}\label{#1}}
\newcommand{\eeadl}[1]{\arrlabel{#1}\end{deqarr}}
\newcommand{\eeal}[1]{\label{#1}\end{eqnarray}}
\newcommand{\eead}[1]{\end{deqarr}}
\newcommand{\eea}{\end{eqnarray}}
\newcommand{\eeaa}{\end{eqnarray*}}

\newcommand{\be}{\begin{equation}}
\newcommand{\ee}{\end{equation}}

\DeclareFontFamily{OT1}{rsfs}{}
\DeclareFontShape{OT1}{rsfs}{m}{n}{ <-7> rsfs5 <7-10> rsfs7 <10->
rsfs10}{} \DeclareMathAlphabet{\mycal}{OT1}{rsfs}{m}{n}

\newcommand{\mcL}{{\mycal L}}
\newcounter{mnotecount}[section]

\newcommand{\rmnote}[1]{}

\def\mysavedown#1{\edef\mysubs{\mysubs#1}}
\def\mysaveup#1{\edef\mysups{\mysups#1}}
\def\mydown#1{{\mytensor}_{\vphantom{\mysubs}#1}}
\def\myup#1{{\mytensor}^{\vphantom{\mysups}#1}}
\def\tensor#1#2{
  #1
  \def\mytensor{\vphantom{#1}}
  \def\mysubs{\relax}
  \def\mysups{\relax}
  \let\down=\mysavedown
  \let\up=\mysaveup
  #2
  \let\down=\mydown
  \let\up=\myup
  #2
  }

\renewcommand{\to}{\rightarrow}

\renewcommand{\epsilon}{\varepsilon}
\renewcommand{\hat}{\widehat}

\def\crn#1#2{{\vcenter{\vbox{
        \hbox{\kern#2pt \vrule width.#2pt height#1pt
           }
          \hrule height.#2pt}}}}

\renewcommand{\hbar}{{\overline h}}

\newcommand{\pre}[2]{{{\vphantom{#2}}^{#1}}\kern-.2ex{#2}}

\theoremstyle{plain}
\newtheorem{theorem}{\sc Theorem}[section]

\newtheorem{bigtheorem} {\sc Theorem}[section]

\newtheorem{Conjecture}[bigtheorem] {\sc Conjecture}

\theoremstyle{definition}

\numberwithin{equation}{section}

\date{\today}

\begin{document}
	\title[Birkhoff-type theorems in linearized gravity]{Non-degeneracy of Riemannian Schwarzschild-anti de Sitter metrics: Birkhoff-type results in linearized gravity}
	\thanks{Preprint UWThPh-2018-18}
	\author{Paul Klinger}
	\address{Paul Klinger,  Faculty of Physics and Erwin Schr\"odinger Institute, University of Vienna, Boltzmanngasse 5, A1090 Wien, Austria}
	\email{paul.klinger@univie.ac.at}
	
	\maketitle
	
\begin{abstract}
	We prove Birkhoff-type results showing that $L^2$ solutions of the linearized Einstein equations around Riemannian Kottler (``Schwarzschild-anti de Sitter'') metrics in arbitrary dimension and horizon topology, which are not controlled by ``master functions'' are pure gauge. 
	Together with earlier results this implies that the $TT$-gauge-fixed linearized Einstein operator for these metrics is non-degenerate for open ranges of the mass parameter.
\end{abstract}
	
\section{Introduction}

In a recent paper, together with Piotr Chru\'sciel and Erwann Delay, we showed that the linearized Einstein operator at a subset of Riemannian Kottler metrics has no $L^2$ kernel\cite{nondegeneracy}. This was motivated by \cite{ChDelayKlingerBH} which gives, for each metric fulfilling this condition, a large class of new stationary black hole spacetimes.

Here we extend the results of \cite{nondegeneracy} to a wider range of dimensions and horizon geometries. In fact the only thing we have to show is that all $L^2$ solutions of the linearized Einstein equations around Riemannian (generalized) Kottler metrics with negative cosmological constant, which are not controlled by the ``master functions'' of Kodama \& Ishibashi \cite{KodamaIshibashiMaster}, have to be pure gauge (except for the case of the critical mass value for spherical horizon geometry). This corresponds to showing that all solutions of the linearized Einstein equations with certain symmetry have to take a fixed form, i.e. a result similar to the Birkhoff theorem in full gravity (see Section \ref{sbirk} below).

Similar results are contained in \cite[Appendices F--I]{nondegeneracy} for spacetime dimension $n+2=4$ for $K\in\{1,-1\}$ and arbitrary dimension for $K=0$ ($K$ is the (constant) sectional curvature of the horizon). Replacing these with the results proved below extends the conclusions of \cite{nondegeneracy} to the stronger
\begin{bigtheorem}
	\label{T15X17.1}
	Let us denote by $P_L$ the linearization,  at Riemannian  Kottler metrics \eq{21III17.1} with negative cosmological constant, of the $TT$-gauge-fixed Einstein operator. Then:
	
	\begin{enumerate}
		\item $P_L$
		has no $L^2$-kernel in spacetime dimension $n+2=4$  except for spherical black holes with mass parameter
		\bel{9VII17.1+5}
		\mhere= \mhere_c:=\frac{n}{n+1}
		\left(\ell
		\sqrt{\frac{n-1}{n+1}}\right)^{n-1}
		\,.
		\ee
		\item  $P_L$
		has no $L^2$-kernel for open ranges of parameters $\mu\in (\mu_{min}(K),\mu(n))$ for $n>2$, where $\mu(n)>\mu_{min}(K)$ solves a polynomial equation and
		\begin{equation}
		\mu_{min}(K):=\begin{cases}
		0&K\in \{0,1\}\,,\\
		-\frac{1
		}{n+1}\left(\frac{n+1}{\ell^2    (n-1)}\right)^{\frac{1-n}{2
		}}& K=-1\,.
		\end{cases}
		\end{equation}
	\end{enumerate}
\end{bigtheorem}
(In contrast to the result of \cite{nondegeneracy} we do not have to restrict to the case $K=0$ for dimensions $n>2$.)

In \cite{nondegeneracy} it is conjectured that
\begin{Conjecture}
	\label{C15X17.21+}
	$P_L$ has no $L^2$-kernel except if $K=1$ and  $\mhere $ is given by \eq{9VII17.1+5}.
\end{Conjecture}
With our results the only missing part to prove Conjecture \ref{C15X17.21+} is a rigorous justification of the numerical arguments in \cite[Section 3.2]{nondegeneracy}.

As mentioned above, the motivation to study the $L^2$ Kernel of $P_L$ comes from \cite{ChDelayKlingerBH}. Indeed, a trivial $L^2$ kernel of $P_L$ for a Riemannian black hole metric $\zgriem$ implies the existence of infinite dimensional families of non-singular, stationary Lorentzian black hole solutions to the Einstein equations with negative cosmological constant, in vacuum or with various matter fields, and with conformal infinity close to that of a Lorentzian metric associated to $\zgriem$.

Theorem \ref{T15X17.1} thus implies the existence of such solutions in all spacetime dimensions and for flat, negatively, or positively curved conformal infinity.

\subsection{The Birkhoff theorem}\label{sbirk}
Our results can be understood as a linearized analogue to the Birkhoff theorem. Several different kinds of results have been referred to as ``Birkhoff theorems'' in the literature (see \cite{Schmidt2013} for an overview). Here we will use the term to mean a classification result showing that under certain symmetry assumptions on a manifold the metric has to take a fixed form (which contains an additional Killing vector field). A classical result of this form is that spherically symmetric vacuum spacetimes are given by the Schwarzschild metric. As far as we are aware the most general such result is \cite[Theorem 3.2]{An:2017wti}. This theorem applies to various kinds of Einstein-matter systems and, in fact, does not even require the full Einstein equations to be satisfied. Specializing to the case of solutions to the vacuum Einstein equations with cosmological constant it states
\begin{theorem}[Birkhoff theorem for warped product vacuum spacetimes \cite{An:2017wti}]\label{t:birkhoff}
	Consider a warped product spacetime $(M=Q\times F, \bar{g}=g+r^2 h)$ satisfying the vacuum Einstein equations with cosmological constant $\Lambda$, where $(Q,g)$ is a 2-dimensional manifold, $(F,h)$ an $n\geq 2$ dimensional one and $r$ is a function on $Q$. Then
	\begin{enumerate}
		\item either $\bar{g}$ takes the standard Eddington-Finkelstein form
		\[
		\bar{g}=-\left(\frac{S^{[h]}}{n(n-1)}-\frac{2m}{r^{n-1}}-\frac{2\Lambda}{n(n+1)}r^2\right)du^2\pm 2 du dr + r^2 h
		\,,
		\]
		where $S^{[h]}=\text{const}$ is the scalar curvature of $h$,
		\item or $\Lambda=0$, $R^{[h]}_{ij}=0$, and
		\[
		\bar{g}=-dt^2+dr^2+ (t\pm r)^2 h
		\,,
		\]
		\item or $r$ is constant, $(Q,g)$ is maximally symmetric, $(F,h)$ is Einstein, $S^{[h]}=2r^2\Lambda$, and $S^{[g]}=4\Lambda/n$.
	\end{enumerate}
\end{theorem}

When $(F, h)$ is $\mathbb{S}^n$ with the round metric this reduces to the classic Birkhoff theorem. In that case (2) does not apply, and (3) gives a limit case of (1) which cannot be described in the standard coordinates (see \cite[Section 4]{Schmidt:1999vp}).

In Section \ref{sl0} we consider perturbations of (Riemannian) Kottler metrics such that, in terms of the variables in Theorem \ref{t:birkhoff}, $\delta h\propto h$ and $\delta g$ is constant on $F$.  We conclude that the only such perturbations which satisfy the linearized Einstein equations are variations of the mass parameter, i.e. ones that (at the linear level) stay in the Kottler family. This is directly analogous to the Birkhoff theorem, with $(F,h)$ being the spaces of constant sectional curvature which appear in the Kottler metrics.

In Section \ref{sl1} we consider axially symmetric perturbations, and conclude that the only ones satisfying the linearized Einstein equations are variations of the angular momentum parameter in the (Riemannian) Kerr anti-de Sitter family. This result is of a similar type as the Birkhoff theorem but has no direct analogue in the nonlinear case.

\section{Definitions \& Background}
We will consider the linearized Einstein equations on a Riemannian (generalized) Kottler~\cite{Kottler} background (also referred to as ``Schwarzschild Anti-de Sitter metrics'' or ``Birmingham metrics''~\cite{Birmingham}). These $n+2$ dimensional solutions of the Einstein equations are given by the manifold
\begin{equation}
M=S^1\times[r_0,\infty)\times {}^n N_K
\end{equation}
where $({}^n N_K,\gamma)$ is an $n$-dimensional space of constant sectional curvature $K\in\{-1,0,1\}$, together with the metric
\bel{21III17.1}
\zgriem = \bigg(\underbrace{\lambdarsquare  +\kappahere  - \frac{2\mhere }{r^{\ourntwo}}}_{=:\VtwoOrF(r)  }\bigg) dt^2
+ \frac{dr^2} {\lambdarsquare  +\kappahere  - \frac{2\mhere }{r^{\ourntwo}}} + r^2 \gamma
\,,
\ee
where $t$ is a periodic coordinate on $S^1$ with period
\[
T:=\frac{f'(r_0)}{4\pi}>0\,,
\]
the parameter $\ell$ is related to the cosmological constant by
$$
\ell  =    \sqrt{-\frac{n(n+1)}{2\Lambda}}  > 0
\,,
$$
and $r_0>0$ is the largest zero of $f$. Note that $r=r_0$ is the axis of rotation for the ``angular'' coordinate $t$.

We use $\mu$,$\nu$,\dots for spacetime indices, $a$,$b$,\dots for indices on $S^1\times[r_0,\infty)$ and $i$,$j$,\dots for those on ${}^nN_K$. We will denote by $\hat{D}_i$, $\hat{\Delta}:=\gamma^{ij}\hat{D}_i\hat{D}_j$ the covariant derivative and Laplace-Beltrami operator on $({}^nN_K, \gamma)$ and by $\tilde{D}_a, \tilde{\Delta}:=\tilde{D}^a\tilde{D}_a$ the corresponding operators on $(S^1\times [r_0,\infty), f dt^2+f^{-1} dr^2)$.

A symmetric $2$-covariant tensor $h$ on $M$ can be split into ``scalar'', ``vector'', and ``tensor'' parts according to their behavior under diffeomorphisms acting on the $n$-dimensional submanifold ${}^n N_K$\cite{Kodama:1985bj}:
\bel{7VI17.101}
\deltazgriem =\deltazgriem ^S+\deltazgriem ^V + \deltazgriem ^T
\,.
\ee

The three parts in \eq{7VI17.101} can be expanded into modes as~\cite[Sections 2.1, 5.1 and 5.2]{KodamaIshibashiMaster}
\begin{align}
	\label{06IX17.1}
	\deltazgriem ^S_{ab}&=\sum_{\efindex} \fabSI \mathbb{S}^{\efindex}\,,
	&\deltazgriem ^S_{ai}&=\sum_{\efindex} r \faSI \mathbb{S}^{\efindex}_i\,,
	&\deltazgriem ^S_{ij}&=\sum_{\efindex} 2r^2(\HLSI\gamma_{ij}\mathbb{S}^{\efindex}+\HTSI\mathbb{S}^{\efindex}_{ij})
	\,,
	\\
	\label{06IX17.2}
	\deltazgriem ^V_{ab}&=0\,,
	&\deltazgriem ^V_{ai}&=\sum_{\efindex} r \faVI \mathbb{V}^{\efindex}_i\,,
	&\deltazgriem ^V_{ij}&=\sum_{\efindex} 2r^2 \HTVI
	\mathbb{V}^{\efindex}_{ij}
	\,,
	\\
	\label{06IX17.3}
	\deltazgriem ^T_{ab}&=0\,,
	&\deltazgriem ^T_{ai}&=0\,,
	&\deltazgriem ^T_{ij}&=\sum_{\efindex} 2r^2 \HTTI
	\mathbb{T}^{\efindex}_{ij}\,,
\end{align}
where the $\mathbb{S}^{\efindex}$, $\mathbb{V}_i^{\efindex}$, $\mathbb{T}_{ij}^{\efindex}$ are scalar, vector, and (symmetric, transverse, and traceless) tensor harmonics, i.e.
\begin{equation}
(\hat{\Delta}_n+k^2)\mathbb{S}^{\efindex}=0\,,
\quad(\hat{\Delta}_n+k_V^2)\mathbb{V}^{\efindex}_i=0\,,\quad
(\hat{\Delta}_n+k_T^2)\mathbb{T}^{\efindex}_{ij}=0\,,
\end{equation}
\begin{equation}
\mathbb{T}^{\efindex}_{ij}=\mathbb{T}^{\efindex}_{ji}\,,\qquad
\gamma^{ij}\hat D_i \mathbb{T}^{\efindex}_{jk}=0\,,\qquad
\gamma^{ij} \mathbb{T}^{\efindex}_{ij}=0\,,
\end{equation}
with eigenvalues $k^2$, $k_V^2$, $k_T^2$ and
\begin{align}
	\mathbb{S}^{\efindex}_i&=-\frac{1}{k}\hat{D}_i\mathbb{S}^{\efindex}
	\,,
	\quad
	k\ne 0
	\,,
	\label{4VII17.26}
	\\
	\label{7IX17.61}
	\mathbb{S}^{\efindex}_{ij}
	&=
	\frac{1}{k^2}\hat{D}_i\hat{D}_j\mathbb{S}^{\efindex}+\frac{1}{n}\gamma_{ij}\mathbb{S}^{\efindex}
	\,,
	\quad
	k \ne 0
	\,,
	\\
	\mathbb{V}^{\efindex}_{ij}&=-\frac{1}{2k_V}(\hat{D}_i\mathbb{V}^{\efindex}_j
	+\hat{D}_j\mathbb{V}^{\efindex}_i)
	=-\frac{1}{2k_V}\mathcal{L}_{\mathbb{V}^{\efindex}}\gamma_{ij}
	\,,
	\quad
	k_V \ne 0
	\,,
\end{align}
with the corresponding quantities vanishing if $k=0$ or $k_V=0$. For the case $K=1$ the eigenvalues are  \cite{Rubin1984}
\begin{align}
	k^2&=l (l+n-1)\,, & l&=0,1,2,\dots\,,\label{K1k}\\
	k_V^2&=l (l+n-1)-1\,, & l&=1,2,3\dots\,,\label{K1kV}\\
	k_T^2&=l(l+n-1)-2\,, & l&=2,3,4,\dots\,, \quad n>2\label{K1kT}\,.
\end{align}

By \cite[Appendix B]{Kodama:1985bj}, using the fact that $({}^nN_K,\gamma)$ is a space of constant curvature, the scalar, vector, and tensor parts of a solution to the linearized Einstein equations separately satisfy the equations.

Kodama and Ishibashi \cite{KodamaIshibashiMaster} introduced master functions, scalar functions $\Phi_{i,\modeindex}$ on the $t,r$ space, satisfying
\bel{7IV16.1}
\tilde{\Delta} {\Phi}_{i,\modeindex  } - V_{i,\modeindex } {\Phi}_{i,\modeindex  }  =0
\,,
\quad
i\in \{S,V,T\}
\,,
\ee
where the $V_{i,\modeindex}(r)$ are some complicated potentials given in \cite[p. 8, 13, 14]{KodamaIshibashiMaster}. These master functions control the behavior of perturbations for all modes for which they are defined. In \cite[Section 3 \& 4]{nondegeneracy} it is shown that whenever the master functions are defined they can be used to prove that there are no $L^2$ solutions of the linearized Einstein equations.

The remaining cases, which have to be treated separately, are
\begin{enumerate}
	\item the $l=0$ scalar and vector modes, i.e. those where $k=0$ or $k_V=0$,
	\item the $l=1$ scalar and vector modes for $K=1$.
\end{enumerate}

We show in the following that $L^2$ perturbations of this form are purely gauge. The first case will be treated in Section \ref{sl0} and the second one in Section \ref{sl1}.

For further reference we note that gauge transformation $h_{\mu\nu}\to h_{\mu\nu}+\mathcal{L}_Y \zgriem_{\mu\nu}$, of perturbations $h$, with (small) gauge vector $Y$, take the form
\beal{18VIII17.1}
h_{tt}&\to&h_{tt}+ Y^r \partial_r \VtwoOrF +2 \VtwoOrF  \partial_t Y^t
\,,
\\
\label{27VIII17.1}
h_{tr}&\to&h_{tr}+ \VtwoOrF^{-1} \partial_t Y^r+\VtwoOrF  \partial_r Y^t
\,,
\\
\label{27VIII17.2}
h_{rr}&\to&h_{rr}+ Y^r \partial_r \VtwoOrF^{-1}+2\VtwoOrF^{-1} \partial_r Y^r
\,,
\\
\label{04IX17.1}
h_{ti}&\to&h_{ti}+ \VtwoOrF \partial_i Y^t+r^2\gamma_{ki}\partial_t Y^k
\,,
\\
\label{04IX17.3}
h_{ri}&\to&h_{ri}+ r^2 \gamma_{ik} \partial_r Y^k + \VtwoOrF^{-1}\partial_i Y^r\,,
\\
\label{27VIII17.3}
h_{ij}&\to&h_{ij} + 2 r Y^r \gamma_{ij} + r^2 (\hat{D}_i(\gamma_{jk}Y^k)+\hat{D}_j(\gamma_{ik}Y^k))
\,.
\eea

By \cite[Proposition 6.5 and Proposition E]{Lee:fredholm} elements of the $L^2$ kernel of $P_L=\Delta_L+2(n+1)$ (see \cite[Section 2]{nondegeneracy}) behave as $|h|_\zgriem=O(r^{-n-1})$ for $r\to\infty$ which gives for the components $h_{\mu\nu}$
\begin{equation}
\begin{aligned}\label{20IX17.3}
h_{tt}&=O(r^{1-n})\,, & h_{tr}&=O(r^{-1-n})\,, &  h_{rr}&=O(r^{-3-n})\,,\\
h_{tj}&=O(r^{1-n})\,, & h_{rj}&=O(r^{-1-n})\,, & h_{jk}&=O(r^{1-n})\,.
\end{aligned}
\end{equation}

\section{The $l=0$ modes for $K\in\{-1,0,1\}$}
\label{sl0}

In this section we show that $L^2$ solutions of the linearized Einstein equations consisting only of $l=0$ modes have to be pure gauge.

For the cases $K=1$ and $K=-1$ we only have to consider the scalar part: The tensor part is always controlled by the master functions and there are no (non-zero) harmonic vectors (i.e. vectors with $k_V=0$) for $K\in\{1,-1\}$. For $K=1$ this can be read of directly from \eqref{K1kV}. For $K=-1$ we consider the Hodge Laplacian
\[
\hat{\Delta}_H \mathbb{V}_i:=(dd^*\mathbb{V}+d^\star d\mathbb{V})_i=-\hat{\Delta}\mathbb{V}_i + {}^n R_i{}^j \mathbb{V}_j\,,
\]
(see e.g. \cite{jost:geomanalysis}).
Using the fact that $\hat{\Delta}_H$ is non-negative and that ${}^n R_{ij}X^iX^j=(n-1)K$ for all unit vectors $X$ (as $({}^nN_{K},\gamma)$ has constant curvature) we obtain, for $K=-1$, $k_V^2 \geq n-1>0$.

\subsection{Scalar perturbations}
We consider the scalar part of a $\ourellnormal=0$  linearized solution $h_{\mu\nu}$  of the Einstein equations, i.e.
\bel{2IX17.1}
h_{ab}= h_{ab}(t,r)
\,,
\quad
h_{ia} \equiv 0
\,,
\quad
h_{ij} = \fpsi (t,r) \zgriem _{ij}=\fpsi(t,r) r^2 \gamma_{ij}
\,,
\ee
and assume that $h\in L^2$.

The angular part of the perturbation can be gauged away by defining a gauge vector $Y$ as
\begin{equation}\label{2IX17.6}
Y^r =  r \fpsi   /2  = O(r^{-n})
\,,
\quad
Y^i \equiv 0
\,,
\end{equation}
which implies
$$
h_{ij} = \mcL_Y \zgriem_{ij}
\,.
$$
The remaining component $Y^t$ of the gauge vector allows us to do the same for $h_{tr}$, by integrating \eq{27VIII17.1} in $r$. However, it is not a priori clear that the resulting gauge vector is smooth at $r=r_0$. We circumvent this problem by cutting off at a finite distance $\epsilon$ from $r_0$, i.e. by defining a gauge vector $Y_\epsilon$ as $Y_\epsilon^r=Y^r$, and
\begin{equation}\label{2IX17.2}
Y_\epsilon^t =   -  \chi_\epsilon(r)\int_{r }^\infty
\VtwoOrF^{- 1}(h_{tr} -  \partial_t Y^r  \VtwoOrF^{-1}) dr
= O(r^{-n-2})
\,,
\end{equation}
where $\chi_\epsilon$ is a smooth function such that $\chi_\epsilon\equiv 1$ for $r>r_0+\epsilon$ and $\chi_\epsilon\equiv 0$ for $r<r_0+\epsilon/2$. With this definition we have, for $r>r_0+\epsilon$,
$$
h_{tr} = \mcL_{Y_\epsilon} \zgriem_{tr}
\,.
$$

We set
\begin{equation}\label{2IX17.3}
\bar h_{\mu\nu} = h_{\mu\nu} - \mcL_{Y_\epsilon} \zgriem_{\mu\nu}
\,,
\end{equation}
thus $\bar h_{\mu\nu}$ is a solution of the linearized Einstein equations with, for $r>r_0+\epsilon$, all components vanishing except possibly $\bar h_{tt}$ and $\bar h_{rr}$.

We now define new functions $Z_r$ and $Z_t$ as
\begin{equation}\label{2IX17.5}
Z_r:= r^{n-1} \VtwoOrF^2 \bar h_{rr}
\,,
\quad
Z_t:= r^{n-1} (\bar h_{tt} + \VtwoOrF^2 \bar h_{rr})
\,,
\end{equation}
chosen such that a variation of the mass in the coordinates of \eq{21III17.1}, which takes the form
\begin{equation}\label{2IX17.4}
2\frac{\delta \mu }{r^{n-1}} ( - dt^2 +  \VtwoOrF^{-2} dr^2 )
\,,
\end{equation}
is captured purely by $Z_r$.

Using \cite[Appendix B]{KodamaIshibashiSeto} we can write the linearized Einstein equations for our perturbation in terms of $Z_r$ and $Z_t$.

For the $t,r$ equation we find, for $r>r_0+\epsilon$,
\begin{equation}\label{2IX17.8}
G_{tr}'[h] = \frac{n \partial_t Z_r(t,r)}{2 \left(-2 \mu r+r^n+r^{2+n}\right)}
\,,
\end{equation}
thus $Z_r$  depends at most upon $r$. One can now eliminate the second radial derivative of $Z_t$ between the $G_{tt}$ and $G_{rr}$ equations, obtaining, again for $r>r_0+\epsilon$,
\begin{equation}\label{5IX17.1}
\partial_r\left(\frac{Z_t}{r^{n-1} \VtwoOrF  }\right)
=0
\,.
\end{equation}
Hence, for $r>r_0+\epsilon$,
\begin{equation}\label{2IX17.9}
Z_t = C(t) r^{n-1}\VtwoOrF
\,.
\end{equation}
for some function $C$ depending only upon $t$.
Inserting all this into the $G_{ij}=0$ equations gives, for $r$ as before, $\partial_r Z_r=0$, and thus $Z_r$ is a constant, say $2\delta \mu$ there.

In terms of $\bar{h}_{rr}$ and $\bar{h}_{tt}$ we now have, for $r>r_0+\epsilon$,
\begin{equation}\label{4IV18.1}
\bar{h}_{tt}=f C(t) -\frac{2\delta\mu}{r}\,,\quad \bar{h}_{rr}=\frac{2\delta\mu}{r f^2}\,.
\end{equation}

As $f$ behaves asymptotically like $r^2$, $C$ has to vanish for this to be in $L^2$.

We find that the only scalar $l=0$ perturbations which satisfy the linearized Einstein equations are, up to gauge, variations of the mass.

For $K\in\{0,-1\}$ the tensor field $\bar h_{\mu\nu}$ is in $L^2$ if and only if $\delta \mu = 0$, while for $K=1$ this holds with the exception of the case $\mu=\mu_c$, with the critical mass $\mu_c$ defined in \eqref{9VII17.1+5}. (See e.g. \cite[Section 2]{nondegeneracy} for a derivation of the critical mass.)

Hence, for these cases, $\bar h_{\mu\nu} \equiv 0$, i.e. $h_{\mu\nu}=\mcL_{Y_\epsilon}\zgriem_{\mu\nu}$, for $r>r_0+\epsilon$. As $\epsilon>0$ is arbitrary and $Y_\delta\equiv Y_\epsilon$ for $r>r_0+\delta,\; \delta>\epsilon$ this applies for all $r>r_0$ with $Y_0$.

The tensors $h$ and $\zgriem$ are smooth by assumption, so we can conclude from $h_{\mu\nu}=\mcL_{Y_0}\zgriem_{\mu\nu}$ that the integrand in \eqref{2IX17.2} is smooth and bounded, implying that $Y:=Y_0$ is in fact smooth for all $r$, including the rotation axis $r=r_0$.

We find that, except for the case of critical mass,
\begin{equation}\label{4IX17.41}
h_{\mu\nu}= \mcL_Y \zgriem_{\mu\nu}
\,,
\qquad
|Y|_{\zgriem} = O(r^{-n-1})
\,,
\end{equation}
i.e. $h$ is pure gauge.
\subsection{Vector perturbations}
For the case $\kappahere=0$ there are (constant) harmonic vectors with $k_V=0$. Perturbations associated with these take the form
\begin{equation}
h_{ab}=0\,,\quad h_{ai}=r f^{V}_a \mathbb{V}_i\,,\quad h_{ij}=0\,,
\end{equation}
where the $\mathbb{V}_i$ are constants and the $f_a^V$ are functions of $t$ and $r$.
Defining $\bar{h}^{V}$ by $h^{V}_{\mu\nu}=\bar{h}^{V}_{\mu\nu}+\mathcal{L}_Y  {} \zgriem$ with a gauge vector $Y$ chosen as $Y^a\equiv 0$ and
\bel{08IX17.2}
Y^ i=
\gamma^{ij}\mathbb{V}_i \int \frac{f^V_{r}}{r} dr
=O(r^{-n-2})\,,
\ee
we obtain $\bar{h}_{ri}=0$, i.e.\ $f_r^V=0$. The removed gauge part behaves asymptotically as
\[
|\mathcal{L}_Y  {}\zgriem|^2_{\zgriem}=O(r^{-2n-2})\,.
\]
We find from \eqref{08IX17.2} that $Y^i$ is regular at $r_0$, and therefore the term $f^{-1}\partial_t Y^i$ which occurs in $|\mathcal{L}_Y  {}\zgriem|_{\zgriem}$ is as well (because of the behavior of $g_{tt}$ there). This implies that $|\mathcal{L}_Y  {}\zgriem|^2_{\zgriem}$ is in $L^2$.

Inserting $\bar{h}$ into the $r,i$ component of the linearized Einstein equations gives
\[
\frac{ r^{n} \left(r \partial_t\partial_r f_t^V-\partial_t f_t^V\right)}{2 \left(r^{n+2}-2 r\mu\right)}=0\,,
\]
and therefore $f_t^V=f_t^V(r)$ by the periodicity of $t$. Inserting back into the $t,i$ equation we obtain
\[
r^2 \partial_r^2 f_t^V + r n \partial_r f_t^V - n f_t^V=0\,,
\]
which gives, after integrating,
\[
f^V_t=r C_1+r^{-n} C_2\,.
\]
Here $C_1$ has to vanish for the perturbation to be in $L^2$ and $C_2$ has to vanish as the tensors $dt dx^i$ are not smooth at the axis of rotation $r=r_0$.
\subsection{Tensor perturbations}
Additionally, for the case $\kappahere=0$ there are (constant) harmonic tensors with $k_T=0$. These are actually controlled by the master functions, but for completeness we show directly that they must vanish.

The associated tensor perturbations take the form
\begin{equation}
h_{ab}=0\,,\quad h_{ai}=0\,,\quad h_{ij}=2 r^2 H_T^T \mathbb{T}_{ij}\,,
\end{equation}
where $\mathbb{T}_{ij}$ is a constant tensor satisfying $\mathbb{T}_{ij}\gamma^{ij}=\mathbb{T}_{ij}\delta^{ij}=0$ and $H_T^T$ is a function of $t$ and $r$ only.

The only nontrivial linearized Einstein equation is
\begin{equation}
\tilde\Delta H^T_T + \frac{n \VtwoOrF}{r}\partial_r H^T_T=0\,.
\end{equation}
This gives $H^T_T=0$ by the maximum principle, as  $H^T_T=O(r^{-1-n})$ from \eqref{20IX17.3}.

\section{The $l=1$ modes for $K=1$}
\label{sl1}

For $K\in\{0,-1\}$ the $l=0$ scalar and vector modes are the only ones not controlled by the master functions of Kodama \& Ishibashi. For $K=1$ however the $l=1$ scalar and vector modes also need to be treated separately. In this section we therefore analyze these $\ourellnormal=1$ modes when $(\Nman,\gamma_{ij})$ is an n-dimensional round unit sphere.  We use the equations of~\cite[Appendix B]{KodamaIshibashiSeto} and our argument is similar to that of~\cite{Dotti2016} in the 2 dimensional case.

\subsection{Vector perturbations}
\label{ss11IX17.1}

The $\ourellnormal=1$ vector perturbations take the form
\bel{08IX17.1}
h^{V}_{\mu\nu}=\sum_{m=1}^{n(n+1)/2}
\begin{pmatrix}
	0 &
	r f^{V}_{a,m} J^{m}_i\\
	r f^{V}_{a,m} J^{m}_i&
	0
\end{pmatrix}\,,
\ee
where the $f_{a,m}^V$ are functions of $t$ and $r$ and $\gamma^{ij} J^{m}_i\partial_j$ form a basis of Killing vector fields on $S^n$ .

Gauge transformations defined by a gauge vector $Y$ of the form
\[
Y^a=0\,,\quad Y^i=\sum_m Y_{m}(t,r)  \gamma^{ij}J^m_j
\,,
\]
preserve the form \eqref{08IX17.1} of the perturbations. The effect of such a gauge transformation on the perturbation is given by
\[
h^{V}_{ai}\to
\sum_m (r f_{a,m}^V + r^2  \partial_a Y_m) J_i^m\,,
\]
with all other components unaffected.

Defining $\bar{h}^{V}$ by $h^{V}_{\mu\nu}=\bar{h}^{V}_{\mu\nu}+\mathcal{L}_Y  {} \zgriem$ with a gauge vector $Y$ given by $Y^ a=0$ and
\bel{08IX17.2}
Y^ i=
\sum_{m} \gamma^{ij}J^m_j \int\limits_{r_0}^r \frac{f^V_{r,m}(t,r')}{r'} dr'
=O(r^{-n-2})\,,
\ee
we find that the components $\bar{h}^{V}_{ri}$ vanish, leaving only $\bar{h}^{V}_{ti}$. The norm of the gauge part is found to be
\[
|\mathcal{L}_Y\zgriem|^2_{\zgriem}=O(r^{-2n-2})\,,
\]
as before, and, as it is regular at $r_0$, $\mathcal{L}_Y  {}\zgriem\in L^2$.

Inserting $\bar h$ into the $r,i$ component of the linearized Einstein equations gives
\[
\frac{r^n \left(r \partial_t\partial_r f^V_t - \partial_t f^V_t\right)}{2 \left(-2 \mu r+r^{n+2}+r^n\right)}=0\,.
\]
Integrating twice and using the periodicity of $t$ we obtain
\[
f^V_t=f^V_t(r)\,.
\]
Inserting into the $t,i$ equation gives
\[
-n f^V_t + n r \partial_r f^V_t + r^2 \partial_r^2 f^V_t=0\,,
\]
and therefore
\[
f^V_t=r C_1+r^{-n} C_2\,,
\]
for constants $C_1$ and $C_2$.
As the tensors $dt dx^i$ are not smooth at the axis of rotation $r=r_0$ we require $C_1 r_0+C_2 r_0^{-n}=0$, i.e.
\[
f^V_t=C_1\frac{r^{n+1}-r_0^{n+1}}{r^n}\,.
\]
Perturbations of this form are exactly variations of the angular momentum parameter $a$ in the Riemannian Kerr anti-de Sitter family (cf. \cite[Appendix J]{nondegeneracy}).

As they are not in $L^2$ we have $\bar{h}=0$ and
\begin{equation}\label{10IX17.12}
h^{V} = \mcL_{Y} \zgriem
\,,
\quad
|Y|_\zgriem = O(r^{-n-1})
\,.
\end{equation}

\subsection{Scalar perturbations}
\label{ss11IX17.2}

Scalar $\ourellnormal=1$ solutions of the linearized Einstein equations take the form
\begin{equation} \label{pert+}
(h_{\alpha \beta}^{S} ) = \sum_m\left( \begin{array}{cc}
f^S_{ab,m} \mathbb{S}^m
&
r f_{a,m}^{S} \mathbb{S}_i^m
\\
r f_{a,m}^{S} \mathbb{S}_i^m
&
2r^2 \mathbb{S}^m H_{L,m}^S \gamma_{ij}
\end{array} \right),
\end{equation}
where $\mathbb{S}_i^m=-k^{-1}\hat D_i \mathbb{S}^m=-n^{-1/2}\hat D_i \mathbb{S}^m$ and the $\mathbb{S}^m$ are the $\ourellnormal=1$ scalar harmonics on $S^n$.

Under gauge transformations
with gauge-vector $Y$ of the form
\begin{equation} \label{vec+}
(Y_\alpha ) = (Y_a, r^2\; \hat D_i X)=\sum_m (\tilde{Y}_{a,m}\mathbb{S}^m, r^2 \tilde{X}_{m} \hat D_i \mathbb{S}^m)
\,,
\end{equation}
where $\tilde{Y}_{a,m}$ and $\tilde{X}_m$ are functions of $t$ and $r$ only, $(h_{\alpha \beta}^{S} )$ transforms to $(\bar{h}_{\alpha \beta}^{S} )$ given by
\begin{equation} \label{pert+gauge+1}
\sum_{m}\left( \begin{array}{cc}
(f^S_{ab,m}+\tilde D_a \tilde{Y}_{b,m}+\tilde D_b \tilde Y_{a,m})\mathbb{S}^m&
(r f^S_{a,m}-\sqrt{n}(\tilde{Y}_{a,m}+r^2\partial_a \tilde X_m))\mathbb{S}^m_i\\
(r f^S_{a,m}-\sqrt{n}(\tilde{Y}_{a,m}+r^2\partial_a \tilde X_m))\mathbb{S}^m_i&
(2r^2 H_L^S+2 \VtwoOrF r \tilde{Y}_{r,m}-2r^2\tilde{X}_m)\mathbb{S}^m\gamma_{ij}
\end{array}
\right)
\,.
\end{equation}

We can use the gauge freedom to set $\bar{h}^S_{ai}=0$ and $\zgriem^{ab} \bar{h}^{S}_{ab}=0$ by choosing $(X,Y_a)$ such that they solve the following system of equations:
\begin{eqnarray}
& \sqrt{n}(\tilde Y_{a,m}+r^2 \partial_a \tilde X_{m})=r f^S_{a,m}=O(r^{-n+1})
\,,&
\label{11IX17.11}
\\
&
\tilde D^b \tilde{Y}_{b,m}=-\frac{1}{2}\zgriem^{ab}f^S_{ab,m}=O(r^{-n-1})
\,.
&
\label{11IX17.13}
\end{eqnarray}

With this choice, $\bar{h}^{S}$ satisfies
\bel{19IX17.07}
\bar{h}^{S}_{ai}=0\,,\qquad
\zgriem^{ab}\bar{h}^{S}_{ab}=0\,.
\ee
Note that \eqref{11IX17.11}-\eqref{11IX17.13} imply
\begin{equation}\label{11IX17.14}
\tilde D^b(r^2 \tilde D_b\tilde{X}_m)=\frac12\zgriem^{ab}f^S_{ab,m}+\frac{1}{\sqrt{n}}\tilde D^b(r f^S_{b,m})
\,.
\end{equation}
The homogeneous version of the equation \eqref{11IX17.14} for $\tilde X_m$ has no non-trivial solutions tending to zero at infinity by the maximum principle. The operator at the left-hand side of \eqref{11IX17.14} has indicial exponents in $\{0,-3\}$, and therefore \eqref{11IX17.14} has a unique solution $\tilde{X}_m = O(r^{-3})$.

The conditions \eqref{19IX17.07} do not fix the gauge uniquely: an additional gauge transformation satisfying
\bel{19IX17.01}
\tilde Y_{a,m}+r^2 \partial_a \tilde X_{m}=0\,,\quad \tilde D^b \tilde{Y}_{b,m}=0\,,
\ee
preserves the form of $\bar{h}^{S}$.

We define new variables $Z_{a,m}$ as
\begin{align}\label{tt1s}
	\begin{split}
		f^S_{tr,m}&=\frac{1}{\VtwoOrF}\Big[Z_{t,m}+
		2r^{1- n}\left(r^n-\mu (n+1) r\right) \partial_t H^S_{L,m}\\
		&\phantom{=\frac{1}{\VtwoOrF}\Big[ }
		-2 r^{4-2 n} \mu (n+1)  \left(-2 \mu +r^{n+1}+r^{n-1}\right) \partial_t\partial_r Z_{r,m}\Big]\,,
	\end{split}\\
	f^S_{rr,m}&=\frac{1}{\VtwoOrF}\Big[2 r \partial_r H^S_{L,m}+2 \mu (n+1) r^{2-n} \partial_r Z_{r,m}\Big]\,.
\end{align}
Note that this defines $Z_r$ only up to a term which depends on $t$ alone.

The $t,r$ linearized Einstein equation directly gives $Z_{t,m}=0$. Eliminating third order derivatives from the remaining equations we obtain
\bel{18IX17.4}\begin{split}
	r^n \partial_r H^S_{L,m}+ r \left(-2 \mu r+r^{n+2}+r^n\right) \partial_r^2 Z_{r,m}&\\
	+ \left(2 \mu (n-2) r+\left(3 r^2+2\right) r^n\right) \partial_r Z_{r,m}&
	=0\,.
\end{split}
\ee
Differentiating the Einstein equations by $r$ and using \eqref{18IX17.4} to express derivatives of $H_{L,m}^S$ by $Z_r^{(1)}$ gives two fifth order and two fourth order equation for $Z_r^{(1)}$. Eliminating higher derivatives we finally obtain a third order equation for $Z_r^{(1)}$
\bel{19IX17.05}\begin{split}
	\partial_r \tilde D^a(r^2 \tilde D_a Z_{r,m})  -2\frac{r^{n-1}-\mu (n+1)}{r^{n+2}+r^n-2 \mu r} \tilde D^a(r^2 \tilde D_a Z_{r,m})=0\,.
\end{split}\ee
This implies
\bel{19IX17.06}
\tilde D^a(r^2 \tilde D_a Z_{r,m})=\frac{Cr^2}{\VtwoOrF}\,,
\ee
with a constant $C$ which has to vanish for $Z_{r,m}$ to be regular at $r_0$.

We now consider the remaining gauge freedom. We see from \eqref{19IX17.01} that for any $X$ satisfying
\bel{19IX17.03}
\tilde D^a(r^2 \tilde D_a \tilde X_m)=0
\ee
there exists an associated $Y_a$ giving a gauge transformation which preserves \eqref{19IX17.07}.

Inserting the definition of our new variables into \eqref{pert+gauge+1} we find that under a gauge transformation satisfying \eqref{19IX17.01} $Z_{r,m}$ and $H_{L,m}^S$ transform as
\beal{19IX17.02}
\partial_r Z_{r,m} &\mapsto& \partial_r( Z_{r,m}+\tilde X_m)
\,,
\\
H^S_{L,m} &\mapsto&  H^S_{L,m} - \tilde{X}_m+\frac{\VtwoOrF}{r}\tilde Y_r\,.
\eea

As, by \eqref{19IX17.06}, $Z_{r,m}$ satisfies \eqref{19IX17.03} we can set $\partial_r Z_{r,m}\equiv 0$ using a gauge transformation with $\partial_r \tilde X_m=-\partial_r Z_{r,m}$, which preserves \eqref{19IX17.07}.

Inserting this into \eqref{18IX17.4} we see that $H_{L,m}^S$ can only depend on $t$. From the remaining equations $\partial_t^2 H_{L,m}^S=0$, i.e. $H_{L,m}^S$ is constant by periodicity.

We can exploit the remaining freedom in $X$ to set
\bel{20IX17.1}
\tilde X_m=H_{L,m}^S\,,\qquad Y_a=0\,,
\ee
obtaining $H_{L,m}^S\equiv 0$.
This gives $Z_{r,m}=\text{const}$ and therefore $f_{tt,m}^S\equiv f_{rr,m}^S\equiv f_{tr,m}^S\equiv 0$.

We arrive at $h^{S}=\mathcal{L}_{\bar{Y}}\zgriem$ where $\bar{Y}$ is the combined gauge vector consisting of the part defined by \eqref{11IX17.11}--\eqref{11IX17.13}, that given by \eqref{19IX17.02} and that given by \eqref{20IX17.1}. From the asymptotics \eqref{20IX17.3} of $h$ and  from \eqref{18VIII17.1}--\eqref{04IX17.3},
with the right-hand sides set to zero,
we conclude that
\bel{20IX17.4}
h^{S}= \mcL_{\bar{Y}} \zgriem
\,,
\quad
|\bar{Y}|_\zgriem=O(r^{-n+1})
\,.
\ee

	\noindent{\sc Acknowledgements:} PK was supported by a uni:docs grant of the University of Vienna. Useful discussions with Piotr Chru\'sciel and Erwann Delay are acknowledged.
	
\printbibliography
\end{document}